\input{aipcheck}

\documentclass[
    ,final 
] {aipproc}

\layoutstyle{6x9}

             \def\gray{$\gamma$-ray}
\def\grays{$\gamma$-rays}

\def\gsim{\mathrel{\raise.5ex\hbox{$>$}\mkern-14mu
             \lower0.6ex\hbox{$\sim$}}}
\def\lsim{\mathrel{\raise.3ex\hbox{$<$}\mkern-14mu
             \lower0.6ex\hbox{$\sim$}}}

\begin{document}

\title{Gamma-ray and Cosmic-ray Tests of Lorentz Invariance Violation and 
Quantum Gravity Models and Their Implications}


\author{Floyd W. Stecker}{
  address={Astrophysics  Science Division,  NASA Goddard  Space Flight
Center, Greenbelt, MD 20771, USA} }

\begin{abstract}

The  topic of  Lorentz  invariance violation  (LIV)  is a  fundamental
question  in  physics  that   has  taken  on  particular  interest  in
theoretical  explorations  of quantum  gravity  scenarios.  I  discuss
various  $\gamma$-ray  observations  that  give  limits  on  predicted
potential effects  of Lorentz  invariance violation.  Among  these are
spectral  data  from  ground   based  observations  of  the  multi-TeV
$\gamma$-rays from nearby AGN,  {\it INTEGRAL} detections of polarized
soft $\gamma$-rays  from the vicinity  of the Crab pulsar,  {\it Fermi
Gamma Ray  Space Telescope} studies of photon  propagation timing from
$\gamma$-ray bursts, and {\it Auger} data on the spectrum of ultrahigh
energy cosmic rays.  These results  can be used to seriously constrain
or  rule out  some models  involving Planck  scale  physics.  Possible
implications  of these  limits for  quantum gravity  and  Planck scale
physics will be discussed.
\end{abstract}

\maketitle


\section{Introduction}

It  has  been  the  major  goal  of particle  physics  to  discover  a
theoretical framework for unifying  gravity with the other three known
forces, {\it viz.}, electromagnetism,  and the weak and strong nuclear
forces. Such a  theory must be compatible with  quantum theory at very
small scales  corrsponding to very  high energies.  Even  the possibly
less  ambitious goal  of reconciling  general relativity  with quantum
theory has been elusive and may require new concepts to accomplish.

There has been a particular interest in the possibility that a quantum
gravity theories  will lead to  Lorentz invariance violation  (LIV) at
the Planck scale, $\lambda_{Pl} = \sqrt{G\hbar /c^3} \sim 1.6
\times 10^{-35}$ m.  This scale corresponds to a mass (energy) scale
of  $M_{Pl}  =  \hbar  /  (\lambda_{Pl}c)  \sim  1.2  \times  10^{19}$
GeV/c$^2$.   It is  at  the  Planck scale  where  quantum effects  are
expected to  play a  key role in  determining the effective  nature of
space-time  that  emerges  as  general  relativity  in  the  classical
continuum limit.  The idea that  Lorentz invariance (LI) may indeed be
only  approximate has  been  explored  within the  context  of a  wide
variety  of suggested Planck-scale  physics scenarios.   These include
the   concepts   of  deformed   relativity,   loop  quantum   gravity,
non-commutative geometry, spin foam  models, and some string theory (M
theory)  models.   Such theoretical  explorations  and their  possible
consequences, such as  observable modifications in the energy-momentum
dispersion  relations  for  free  particles  and  photons,  have  been
discussed under the general heading of ``Planck scale phenomenology''.
There is  an extensive literature on this  subject.  (See ~\cite{ma05}
for  a  review;  some   recent  references  are  Refs.~\cite{el08}  --
~\cite{he09}.   For a  non-technical  treatment of  the present  basic
approaches to a quantum  gravity theory, see Ref.~\cite{smolin}).  One
should keep  in mind that in  a context that is  separate from quantum
gravity considerations,  it is important to  test LI for  its own sake
~\cite{co98,cg99}.  {\it LIV gratia  LIV}. The significance of such an
approach is  evident when one considers the  unexpected discoveries of
the violation of  $P$ and $CP$ symmetries. In fact,  it has been shown
that a violation of $CPT$ would imply LIV ~\cite{gr02}

We will consider here some of the consequent searches for such effects
using high energy astrophysics observations, particularly observations
of high energy cosmic $\gamma$-rays and ultrahigh energy cosmic rays.

\section{LIV Perturbations}

We know  that Lorentz invariance  has been well validated  in particle
physics; indeed, it plays an essential role in designing machines such
as  the new  LHC (Large  Hadron Collider).   Thus, any  LIV  extant at
accelerator energies (``low energies'') must be extremely small.  This
consideration is reflected by  adding small Lorentz-violating terms in
the  free particle  Lagrangian.  Such  terms can  be postulated  to be
independent    of     quantum    gravity    theory,     {\it    e.g.},
Refs.~\cite{co98,cg99}.   Alternatively, it  can be  assumed  that the
terms are small  because they are suppressed by one  or more powers of
$p/M_{Pl}$ (with  the usual convention that  $c = 1$.)   In the latter
case, in the  context of effective field theory  (EFT), such terms are
assumed  to  approximate  the  effects  of quantum  gravity  at  ``low
energies'' when $p \ll M_{Pl}$.

One result  of such  assumptions is a  modification of  the dispersion
relation that  relates the energy and  momentum of a  free particle or
photon.  This,  in turn,  can lead to  a maximmum  attainable velocity
(MAV) of a particle different from  $c$ or a variation of the velocity
of a photon {\it in vacuo} with photon energy.  Both effects are clear
violations of relativity theory.  Such modifications of kinematics can
result  in changes  in threshold  energies for  particle interactions,
suppression  of  particle interactions  and  decays,  or allowance  of
particle interactions  and decays that are  kinematically forbidden by
Lorentz invariance ~\cite{cg99}.

A  simple  formulation  for  breaking   LI  by  a  small  first  order
perturbation  in  the  electromagnetic  Lagrangian which  leads  to  a
renormalizable   treatment    has   been   given    by   Coleman   and
Glashow~\cite{cg99}.   The small  perturbative noninvariant  terms are
both  rotationally  and   translationally  invariant  in  a  preferred
reference frame  which one  can assume  to be the  frame in  which the
cosmic background  radiation is isotropic. These terms  are also taken
to   be  invariant  under   $SU(3)\otimes  SU(2)\otimes   U(1)$  gauge
transformations in the standard model.

Using  the formalism  of  Ref.~\cite{cg99},  we denote  the  MAV of  a
particle of type  $i$ by $c_{i}$, a quantity  which is not necessarily
equal to  $c \equiv 1$,  the low energy  {\it in vacua\/}  velocity of
light.  We further define the difference $c_{i} - c_{j} \equiv
\delta_{ij}$.  These  definitions can be generalized and can be used to  
discuss the physics implications of cosmic-ray and cosmic $\gamma$-ray
observations~\cite{sg01 -- st09}.

\section{Electroweak Interactions}

In general then, $c_e \ne c_\gamma$. The physical consequences of such
a violation of  LI depend on the sign of  the difference between these
two MAVs. Defining

\begin{equation}
c_{e} \equiv c_{\gamma}(1 + \delta) ~ , ~ ~~~0< |\delta| \ll 1\;,
\label{delta}
\end{equation}

\noindent
one  can consider the  two cases  of positive  and negative  values of
$\delta$ separately~\cite{cg99,sg01}.

{\it Case I:} If $c_e<c_\gamma$ ($\delta  < 0$), the decay of a photon
into an  electron-positron pair  is kinematically allowed  for photons
with energies exceeding

\begin{equation}
E_{\rm max}= m_e\,\sqrt{2/|\delta|}\;.
\end{equation}

\noindent
The  decay would  take place  rapidly, so  that photons  with energies
exceeding $E_{\rm max}$ could not be observed either in the laboratory
or as cosmic rays. From the  fact that photons have been observed with
energies $E_{\gamma} \ge$ 50~TeV from the Crab nebula, one deduces for
this case  that $E_{\rm max}\ge  50\;$TeV, or that -$\delta  < 2\times
10^{-16}$.

{\it Case  II:} For this possibility, where  $c_e>c_\gamma$ ($\delta >
 0$), electrons  become superluminal if their  energies exceed $E_{\rm
 max}/2$.  Electrons  traveling faster than  light will emit  light at
 all frequencies by a process  of `vacuum \v Cerenkov radiation.' This
 process  occurs  rapidly,  so  that  superluminal  electron  energies
 quickly  approach $E_{\rm max}/2$.   However, because  electrons have
 been seen in the cosmic  radiation with energies up to $\sim\,$2~TeV,
 it  follows that $E_{\rm  max} \ge  2$~TeV, which  leads to  an upper
 limit on  $\delta$ for this case  of $3 \times  10^{-14}$.  Note that
 this limit is two orders  of magnitude weaker than the limit obtained
 for  Case I.   However, this  limit can  be considerably  improved by
 considering constraints obtained from  studying the \gray\ spectra of
 active galaxies~\cite{sg01}.

\subsection{Constraints on LIV from AGN Spectra}

A constraint on $\delta$ for $\delta > 0$ follows from a change in the
threshold energy for the pair production process $\gamma + \gamma
\rightarrow e^+ + e^-$.  This follows from the fact that the square of
the four-momentum is changed to give the threshold condition

\begin{equation}
2\epsilon   E_{\gamma}(1-cos\theta)~   -~  2E_{\gamma}^2\delta   ~\ge~
4m_{e}^2,
\label{threshold}
\end{equation}

\noindent where $\epsilon$ is the  energy of the low energy photon and
$\theta$ is the angle between the  two photons. The second term on the
left-hand-side  comes  from  the  fact  that  $c_{\gamma}  =  \partial
E_{\gamma}/\partial p_{\gamma}$.  It follows  that the condition for a
significant increase  in the energy  threshold for pair  production is
$E_{\gamma}\delta/2$  $ \ge$  $ m_{e}^2/E_{\gamma}$,  or equivalently,
$\delta \ge {2m_{e}^{2}/E_{\gamma}^{2}}$.  ~ The observed $\gamma$-ray
spectrum  of the active  galaxies Mkn  501 and  Mkn 421  while flaring
~\cite{ah01}  exhibited  the  high  energy  absorption  expected  from
$\gamma$-ray    annihilation    by    extragalactic    pair-production
interactions  with  extragalactic  infrared  photons~\cite{ds02,ko03}.
This led Stecker and Glashow~\cite{sg01} to point out that the Mkn 501
spectrum presents  evidence for pair-production with  no indication of
LIV up  to a photon  energy of $\sim\,$20~TeV  and to thereby  place a
quantitative    constraint    on    LIV    given    by    $\delta    <
2m_{e}^{2}/E_{\gamma}^{2} \simeq 10^{-15}$.

\section{Gamma-ray Constraints on Quantum Gravity and Extra Dimension Models}

As previously mentioned, LIV has  been proposed to be a consequence of
quantum  gravity physics  at the  Planck scale  ~\cite{ga95,al02}.  In
models  involving large extra  dimensions, the  energy scale  at which
gravity becomes strong  can occur at a quantum  gravity scale, $M_{QG}
<< M_{Pl}$, even approaching  a TeV~\cite{ar98}.  In the most commonly
considered case,  the usual relativistic  dispersion relations between
energy   and   momentum  of   the   photon   and   the  electron   are
modified~\cite{al02,ac98} by a term of order $p^3/M_{QG}$.

Generalizing the LIV parameter $\delta$ from equation (\ref{delta}) to
an energy dependent form, we find

\begin{equation}
\delta~  \equiv~ {\partial  E_{e}\over{\partial p_{e}}}~  -~ {\partial
E_{\gamma}         \over{\partial         p_{\gamma}}}~        \simeq~
{E_{\gamma}\over{M_{QG}}}~      -~{m_{e}^{2}\over{2E_{e}^{2}}}~     -~
{E_{e}\over{M_{QG}}} .
\label{pcube}
\end{equation}

It follows that  the threshold condition for pair  production given by
equation      (\ref{threshold})     implies      that     $M_{QG}~\ge~
E_{\gamma}^3/8m_{e}^2.$ Since  pair production occurs  for energies of
at  least  20  TeV,  we  find  a constraint  on  the  quantum  gravity
scale~\cite{st03}   $M_{QG}   \ge   0.3  M_{Pl}$.    This   constraint
contradicts the  predictions of  some proposed quantum  gravity models
involving large extra dimensions  and smaller effective Planck masses.
In  a  variant  model  of  Ref. ~\cite{el04},  the  photon  dispersion
relation is changed, but not that  of the electrons.  In this case, we
find the even stronger constraint $M_{QG} \ge 0.6 M_{Pl}$.

\section{Energy Dependent Photon Delays from GRBs  and Tests of 
Lorentz Invariance Violation}

One  possible  manifestation  of  Lorentz invariance  violation,  from
Planck scale physics produced by  quantum gravity effects, is a change
in the  energy-momentum dispersion  relation of a  free particle  or a
photon. If  this results from  the linear Planck-supressed term  as in
equation~(\ref{pcube})  above,  this  results  in  a  photon  velocity
retardation       that      is       of      first       order      in
$E_{\gamma}/M_{QG}$~\cite{ac98,el00}.   In a $\Lambda  CDM$ cosmology,
where present observational data indicate that $\Omega_{\Lambda}
\simeq 0.7$  and $\Omega_{m} \simeq 0.3$, the  resulting difference in
the  propagation times  of  two photons  having  an energy  difference
$\Delta E_{\gamma}$ from a $\gamma$-ray  burst (GRB) at a redshift $z$
will be

\begin{equation}
\Delta  t_{LIV}   =  H_{0}^{-1}  {{\Delta   E_{\gamma}}  \over  M_{QG}
}{\int_0^z}{{dz'(1+z')} \over {\sqrt{\Omega_{\Lambda} +
\Omega_{m}(1+z')^3}}}
\label{delay}
\end{equation}

\noindent for  a photon dispersion of  the form $c_{\gamma}  = c(1 -
E_{\gamma}/M_{QG}$), with  $c$ being the usual low  energy velocity of
light~\cite{ja08}. In  other words,  $\delta$, as defined  earlier, is
given by $- E_{\gamma}/M_{QG}$.

\begin{figure}[h]
\includegraphics[height=.25\textheight]{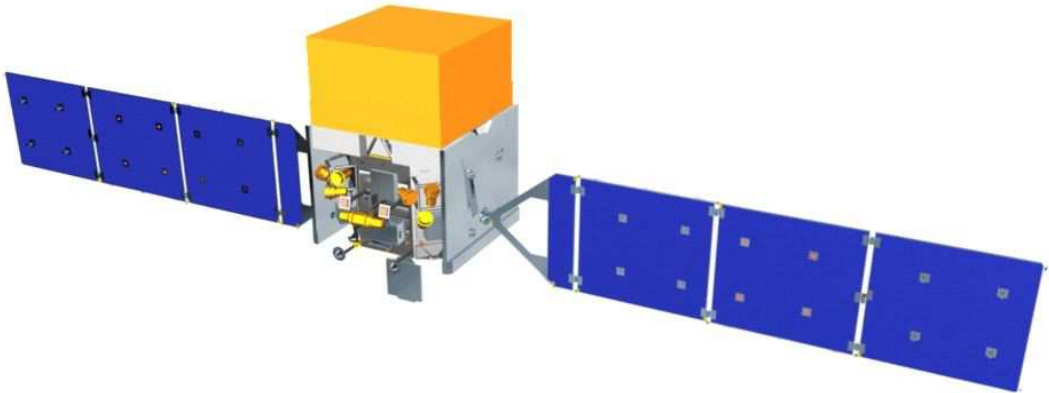}
\caption{Schematic of the {\it Fermi} satellite, launched in June of
2008. The {\it LAT}  is located at the top (yellow  area) and the {\it
GBM} array is located directly below.}
\label{glast}
\end{figure}

The {\it  Fermi} Gamma-ray Space Telescope,  (see Figure \ref{glast}),
with its \gray\ {\it Burst Monitors (GBM)} covers an energy range from
8 keV  to 40 MeV  and its {\it  Large Area Telescope (LAT)}  covers an
energy range from  20 MeV to $> 300$ GeV. \footnote  {See paper the of
Silvia Rain\`{o},  these proceedings.}  It can observe  and study both
GRBs and flares from active galactic nuclei over a large range of both
energy and  distance. This was the  case with the GRB  090510, a short
burst at  a cosmological distance  corresponding to a redshift  of 0.9
that produced photons with energies  extending from the X-ray range to
a \gray\ of  energy $\sim$ 31 GeV. This burst  was therefore a perfect
subject for  the application  of equation (\ref{delay}).   {\it Fermi}
observations  of GRB090510  have yielded  the best  constraint  on any
first order retardation of photon velocity with energy $\Delta t
\propto (E/M_{QG})$. This result would require a value of $M_{QG} \gsim
1.2 M_{Pl}$~\cite{Fermi2009}\footnote {See also the paper of Francesco
de Palma, these proceedings.}  In large extra dimension scenarios, one
can have  effective Planck masses  smaller than $1.22  \times 10^{19}$
GeV, whereas in  most QG scenarios, one expects  that the minimum size
of space-time  quanta to be  $\lambda_{Pl}$. This implies a  value for
$M_{QG} \lsim M_{Pl}$ in all cases.

In  particular,   we  note  the   string  theory  inspired   model  of
Ref.~\cite{el08}.   This  model  invisions  space-time  as  a  gas  of
D-particles in a higher dimensional bulk where the observable universe
is  a D3  brane. The  photon  is represented  as an  open string  that
interacts  with  the  D-particles,  resulting a  retardation  $\propto
E_{\gamma}/M_{QG}$. The new  {\it Fermi} data appear to  rule out this
model as well as other models that predict such a retardation.

The dispersion effect will be smaller if the dispersion relation has a
quadratic dependence on  $E_{\gamma}/M_{QG}$ as suggested by effective
field theory  considerations~\cite{my03,ja04}.  This will  obviate the
limits on $M_{QG}$ given above.  These considerations also lead to the
prediction of vacuum birefringence (see next section).

\section{Looking for Birefringence Effects from Quantum Gravity}

A  possible model for  quantizing space-time  which has  been actively
investigated is  {\it loop quantum  gravity} (see the review  given in
Ref. ~\cite{pe04}  and references therein.) A signature  of this model
is  that the  quantum nature  of space-time  can produce  a vacuum
birefringence    effect.    (See   also    the   EFT    treatment   in
Ref.~\cite{my03}.)  This is  because electromagnetic waves of opposite
circular polarizations will propagate with different velocities, which
leads to a rotation of linear polarization direction through the angle
\begin{equation} 
\theta(t)=\left[\omega_+(k)-\omega_-(k)\right]t/2=\xi               k^2
t/2M_{Pl}
\label{rotation}
\end{equation}
for a  plane wave with wave-vector $k$~\cite{ga99}.  Again, for simple
Planck-suppressed LIV, we would expect that $\xi \simeq 1$.

Some astrophysical sources emit  highly polarized radiation. It can be
seen from equation (\ref{rotation}) that the rotation angle is reduced
by the  large value of the  Planck mass. However,  the small rotations
given  by equation (\ref{rotation})  can add  up over  astronomical or
cosmological  distances  to  erase  the  polarization  of  the  source
emission. Therefore, if  polarization is seen in a  distant source, it
puts constraints  on the  parameter $\xi$.  Observations  of polarized
radiation from distant sources can therefore be used to place an upper
bound on $\xi$.

Equation (\ref{rotation})  indicates that  the higher the  wave number
$|k|$,  the   stronger  the  rotation  effect  will   be.   Thus,  the
depolarizing effect  of space-time induced birefringence  will be most
pronounced in the  \gray\ energy range.  It can also  be seen that the
this effect grows linearly with propoagation time.

The difference in rotation angles for wave-vectors $k_1$ and $k_2$ is

\begin{equation}
 \Delta\theta=\xi (k_2^2-k_1^2) d/2M_{Pl}, \label{diffrotation}
\end{equation}
replacing  the  time  $t$ by  the  distance  from  the source  to  the
detector, denoted by $d$.

The  best  secure bound  on  this  effect,  $|\xi|\lsim 10^{-9}$,  was
obtained using  the observed 10\% polarized soft  \gray\ emission from
the region of the Crab Nebula~\cite{ma08}.

Clearly,  the best  tests of  birefringence  would be  to measure  the
polarization of \grays\ from GRBs. We note that linear polarization in
X-ray flares  from GRBs  has been predicted~\cite{fa05}.   Most \gray\
bursts have redshifts  in the range 1-2 corresponding  to distances of
greater than  a Gpc.  Should polarzation  be detected from  a burst at
distance $d$, this would place a limit on $|\xi|$ of

\begin{equation}
|\xi| \lsim 5 \times 10^{-15}/d_{0.5}
\end{equation}

\noindent where $d_{0.5}$ is the distance to the burst in units of 
0.5 Gpc  ~\cite{ja04}.  Detectors  that are dedicated  to polarization
measurements in  the X-ray  and \gray\ energy  range and which  can be
flown  in space to  study the  polarization from  distant astronomical
sources are now being designed~\cite{mi05,pr05}.

\section{LIV and the Ultrahigh Energy Cosmic Ray Spectrum}

\subsection{The ``GZK Effect''}

Shortly after the discovery  of the 3K cosmogenic background radiation
(CBR),  Greisen  ~\cite{gr66} and  Zatsepin  and Kuz'min  ~\cite{za66}
predicted that pion-producing interactions  of such cosmic ray protons
with the CBR should produce a spectral cutoff at $E \sim$ 50 EeV.  The
flux  of  ultrahigh energy  cosmic  rays  (UHECR)  is expected  to  be
attenuated by such photomeson  producing interactions.  This effect is
generally known as the ``GZK  effect''.  Owing to this effect, protons
with energies above $\sim$100~EeV  should be attenuated from distances
beyond $\sim 100$ Mpc because  they interact with the CBR photons with
a resonant photoproduction of pions ~\cite{st68}.

\subsection{Modification of the GZK Effect Owing to LIV}

Let us consider  the photomeson production process leading  to the GZK
effect. Near threshold, where single pion production dominates,

\begin{equation}
p + \gamma \rightarrow p + \pi.
\end{equation}

Using the  normal Lorentz  invariant kinematics, the  energy threshold
for  photomeson interactions  of UHECR  protons of  initial laboratory
energy $E$ with  low energy photons of the  CBR with laboratory energy
$\omega$, is  determined by the relativistic invariance  of the square
of  the  total  four-momentum   of  the  proton-photon  system.   This
relation, together with the threshold inelasticity relation $E_{\pi} =
m/(M  +  m)  E$  for  single pion  production,  yields  the  threshold
conditions for head on collisions in the laboratory frame

\begin{equation}
4\omega E = m(2M + m)
\end{equation}

\noindent for the proton, and

\begin{equation}
4\omega E_{\pi} = {{m^2(2M + m)} \over {M + m}}
\label{pion}
\end{equation}

\noindent in terms of the pion energy, where M is the rest mass of the
proton and m is the rest mass of the pion~\cite{st68}.

If LI  is broken  so that  $c_\pi~ >~ c_p$,  the threshold  energy for
photomeson   is  altered.\footnote{This   requirement   precludes  the
`quasi-vacuum
\v{C}erenkov  radiation'  of  pions,   {\it  via}  the  rapid,  strong
interaction,  pion emission process,  $p \rightarrow  N +  \pi$.  This
process would be allowed by LIV  in the case where $\delta_{\pi p}$ is
negative, producing a sharp cutoff  in the UHECR proton spectrum. (For
more details, see Refs.~\cite{cg99,st09,alt07}.}

Because  of  the  small  LIV  perturbation term,  the  square  of  the
four-momentum  is shifted  from  its  LI form  so  that the  threshold
condition in terms of the pion energy becomes

\begin{equation}
4\omega E_{\pi}  = {{m^2(2M +  m)} \over {M  + m}} + 2  \delta_{\pi p}
E_{\pi}^2
\label{LIVpi}
\end{equation}

\noindent where $ \delta_{\pi p} \equiv  c_\pi~ - ~ c_p$, again in units
where the low energy velocity of light is unity.

Equation (\ref{LIVpi})  is a quadratic  equation with real  roots only
under the condition

\begin{equation}
\delta_{\pi p}  \le {{2\omega^2(M  + m)} \over  {m^2(2M +  m)}} \simeq
\omega^2/m^2.
\label{root}
\end{equation}

Defining  $\omega_0 \equiv  kT_{CBR} =  2.35 \times  10^{-4}$  eV with
$T_{CBR} = 2.725\pm 0.02$ K, equation (\ref{root}) can be rewritten

\begin{equation}
\delta_{\pi p} \le 3.23 \times 10^{-24} (\omega/\omega_0)^2.
\label{CG}
\end{equation}

\subsection{Kinematics}

If LIV occurs and $\delta_{\pi p} > 0$, photomeson production can only
take place for interactions of  CBR photons with energies large enough
to satisfy equation (\ref{CG}). This condition, together with equation
(\ref{LIVpi}), implies  that while photomeson  interactions leading to
GZK suppression can occur for ``lower energy'' UHE protons interacting
with higher energy CBR photons on the Wien tail of the spectrum, other
interactions involving higher energy  protons and photons with smaller
values  of  $\omega$ will  be  forbidden.   Thus,  the observed  UHECR
spectrum may  exhibit the characteristics of GZK  suppression near the
normal GZK threshold, but the UHECR spectrum can ``recover'' at higher
energies  owing to  the  possibility that  photomeson interactions  at
higher  proton energies  may be  forbidden.   We now  consider a  more
detailed quantitative  treatment of this possibility,  {\it viz.}, GZK
coexisting with LIV.

The  kinematical  relations   governing  photomeson  interactions  are
changed  in  the  presence  of  even  a  small  violation  of  Lorentz
invariance.  The modified kinematical  relations containing LIV have a
strong  effect on  the amount  of  energy transfered  from a  incoming
proton to the pion produced in the subsequent interaction, {\it i.e.},
the inelasticity ~\cite{st09,al03,ss08}.

The primary  effect of LIV on  photopion production is  a reduction of
phase space allowed for the interaction.  This results from the limits
on the allowed range of interaction angles integrated over in order to
obtain   the  total   inelasticity.   For   real-root   solutions  for
interactions   involving   higher  energy   protons,   the  range   of
kinematically  allowed   angles  becomes  severely   restricted.   The
modified  inelasticity that  results  is the  key  in determining  the
effects of LIV on photopion production. The inelasticity rapidly drops
for higher incident proton energies.

\begin{figure}[h]
\includegraphics[height=.3\textheight]{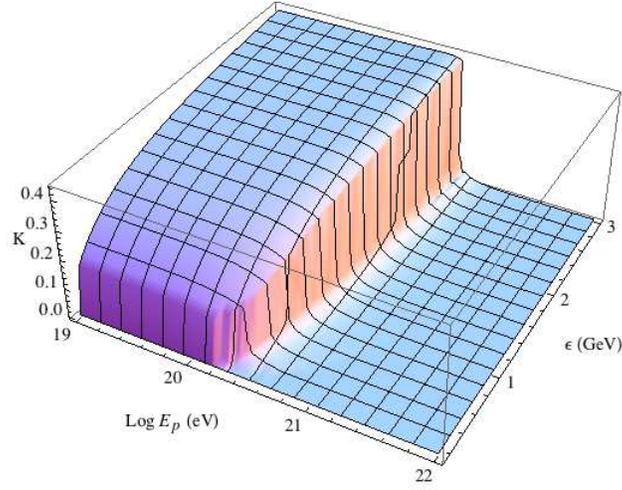}
\caption{The  calculated  proton  inelasticity  modified  by  LIV  for
$\delta_{\pi  p} =  3 \times  10^{-23}$ as  a function  of  CBR photon
energy and proton energy \protect ~\cite{ss08}.}
\label{inelasticity}
\end{figure}

Figure  \ref{inelasticity} shows  the  calculated proton  inelasticity
modified by LIV for a value of $\delta_{\pi p} = 3 \times 10^{-23}$ as
a function of  both CBR photon energy and  proton energy ~\cite{ss08}.
Other choices for $\delta_{\pi p}$ yield similar plots.  The principal
result  of changing the  value of  $\delta_{\pi p}$  is to  change the
energy  at which  LIV effects  become  significant.  For  a choice  of
$\delta_{\pi p}  = 3 \times  10^{-23}$, there is no  observable effect
from LIV for $E_{p}$ less  than $\sim200$ EeV.  Above this energy, the
inelasticity  precipitously drops  as the  LIV term  in the  pion rest
energy approaches $m_{\pi}$.

With  this  modified inelasticity,  the  proton  energy  loss rate  by
photomeson production is given by

\begin{equation}
{{1}\over{E}}{{dE}\over{dt}} = - {{\omega_{0}c}\over{2\pi^2
\gamma^2}\hbar^3c^3}  \int\limits_\eta^\infty d\epsilon  ~  \epsilon ~
\sigma(\epsilon)                                            K(\epsilon)
\ln[1-e^{-\epsilon/2\gamma\omega_{0}}]\end{equation}

\noindent where we  now use $\epsilon$ to designate  the energy of the
photon  in the  cms, $\eta$  is the  photon threshold  energy  for the
interaction in  the cms,  $K(\epsilon)$ denotes the  inelasticity, and
$\sigma(\epsilon)$  is   the  total  $\gamma$-p   cross  section  with
contributions from  direct pion production,  multipion production, and
the $\Delta$ resonance.

\begin{figure}[h]
\includegraphics[height=.3\textheight]{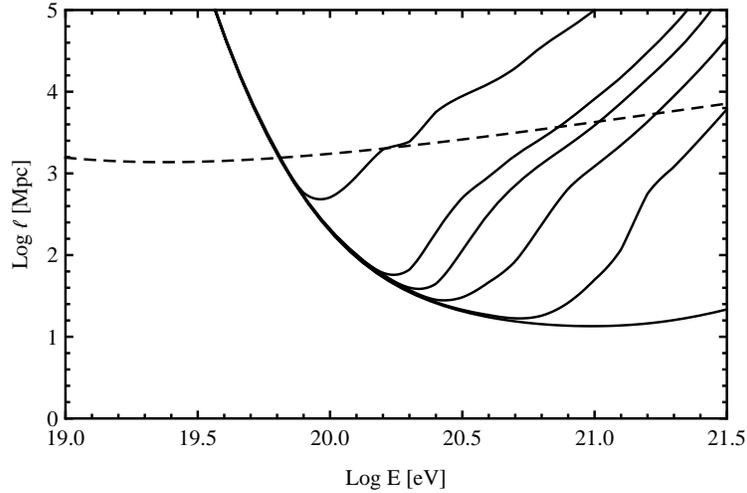}
\caption{The  calculated  proton  attenuation  lengths as  a  function
proton energy modified  by LIV for various values  of $\delta_{\pi p}$
(solid lines),  shown with the attenuation length  for pair production
unmodified by LIV  (dashed lines). From top to  bottom, the curves are
for $\delta_{\pi p} = 1 \times  10^{-22}, 3 \times 10^{-23} , 2 \times
10^{-23},  1  \times  10^{-23},  3  \times 10^{-24},  0$  (no  Lorentz
violation) \protect ~\cite{ss08}.}
\label{attn}
\end{figure}

The  corresponding  proton attenuation  length  is  given  by $\ell  =
cE/r(E)$,  where the  energy loss  rate $r(E)  \equiv  (dE/dt)$.  This
attenuation length is plotted  in Figure \ref{attn} for various values
of  $\delta_{\pi  p}$  along   with  the  unmodified  pair  production
attenuation length from pair production interactions, $p +
\gamma_{CBR} \rightarrow e^+ + e^-$.

\section{UHECR Spectra with LIV and Comparison with Present Observations}

The effect of by a very small  amount of LIV on the UHECR spectrum was
analytically calculated in Ref.~\cite{ss08}  in order to determine the
resulting spectral  modifications.  It can be  demonstrated that there
is  little  difference  between  the  results  of  using  an  analytic
calculation  {\it vs.}   a Monte  Carlo calculation  ({\it  e.g.}, see
Ref. ~\cite{ta09}). In order to  take account of the probable redshift
evolution  of  UHECR production  in  astronomical  sources, they  took
account of the following considerations: \\

\noindent  ({\it  i})  The  CBR  photon number  density  increases  as
$(1+z)^3$ and the CBR  photon energies increase linearly with $(1+z)$.
The corresponding energy loss for  protons at any redshift $z$ is thus
given by

\begin{eqnarray}
r_{\gamma p}(E,z) = (1+z)^3 r[(1+z)E].
\label{eq5}
\end{eqnarray}

\noindent  ({\it  ii})  They assumed  that  the  average  UHECR  volume
emissivity  is of  the energy  and  redshift dependent  form given  by
$q(E_i,z) =  K(z)E_i^{-\Gamma}$ where $E_i$  is the initial  energy of
the  proton  at  the source  and  $\Gamma  =  2.55$.  For  the  source
evolution, we assume $K(z) \propto (1  + z)^{3.6}$ with $z \le 2.5$ so
that  $K(z)$ is  roughly  proportional to  the empirically  determined
$z$-dependence of  the star formation rate. $K(z=0)$  and $\Gamma$ are
normalized fit the data below the GZK energy.

\begin{figure}
\includegraphics[height=.3\textheight]{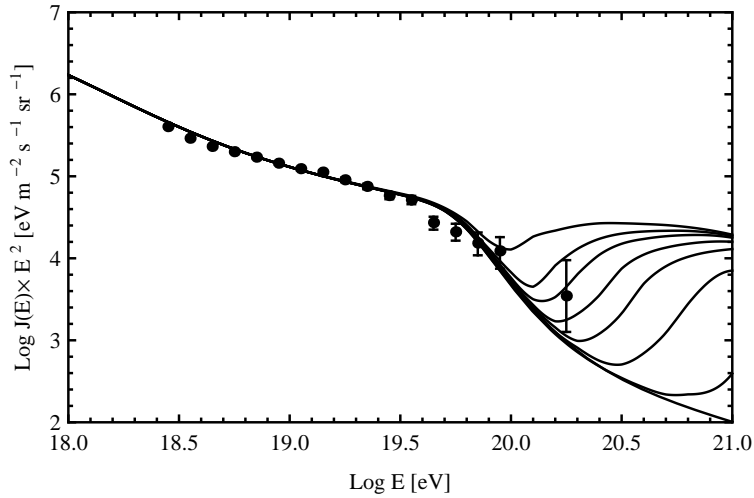}
\caption{Comparison of  the latest Auger data  with calculated spectra
for various  values of  $\delta_{\pi p}$, taking  $\delta_p =  0$ (see
text).  From top to bottom,  the curves give the predicted spectra for
$\delta_{\pi p}  = 1  \times 10^{-22}, 6  \times 10^{-23},  4.5 \times
10^{-23}, 3 \times 10^{-23} , 2 \times 10^{-23}, 1 \times 10^{-23}, 3
\times 10^{-24}, 0$ (no Lorentz violation) \protect ~\cite{st09}.}
\label{Auger}
\end{figure}

Using these  assumptions, one can calculate  the effect of  LIV on the
UHECR spectrum.   The results are actually insensitive  to the assumed
redshift dependence because evolution does not affect the shape of the
UHECR  spectrum  near the  GZK  cutoff  energy ~\cite{be88,st05}.   At
higher energies  where the attenuation  length may again  become large
owing to an  LIV effect, the effect of evolution turns  out to be less
than 10\%.  The curves calculated in Ref.~\cite{st09} assuming various
values of $\delta_{\pi p}$, are shown in Figure \ref{Auger} along with
the latest {\it  Auger} data from Ref. ~\cite{sch09}.   They show that
{\it even  a very small amount of  LIV that is consistent  with both a
GZK effect and with the present  UHECR data can lead to a ``recovery''
of the UHECR spectrum at higher energies.}

\subsection{Allowed Range for the LIV Parameter $\delta_{\pi p}$}

Stecker   and   Scully   ~\cite{st09}   have  updated   compared   the
theoretically predicted  UHECR spectra with various amounts  of LIV to
the  latest  {\it  Auger}  data   from  the  procedings  of  the  2009
International  Cosmic Ray Conference  ~\cite{sch09},~\cite{data}. This
update  is  shown in  Figure  \ref{Auger}.   The  amount of  presently
observed  GZK suppression  in the  UHECR data  is consistent  with the
possible  existence  of   a  small  amount  of  LIV.    The  value  of
$\delta_{\pi p}$ that results in the smallest $\chi^2$ for the modeled
UHECR  spectral fit  using  the observational  data  from {\it  Auger}
~\cite{sch09} above the GZK energy.   The best fit LIV parameter found
was in the range given by $\delta_{\pi p}$ = $3.0^{+1.5}_{-3.0} \times
10^{-23}$, corresponding to an upper limit on $\delta_{\pi p}$ of $4.5
\times 10^{-23}$.
\footnote{The {\it HiRes} data ~\cite{ab08}  do not reach a high enough
energy to  further restrict LIV.}  \footnote{We note  that the overall
fit of  the data  to the theoretically  expected spectrum  is somewhat
imperfect,  even below  the GZK  energy and  even for  the case  of no
LIV. It  appears that the {\it  Auger} spectrum seems  to steepen even
below  the GZK  energy.   As a  conjecture,  one can  assume that  the
derived energy may be too low by about 25\%, within the uncertainty of
both   systematic-plus  statistical   error  given   for   the  energy
determination.   This gives better  agreement between  the theoretical
curves and the shifted data  ~\cite{st09}. The constraint on LIV would
be only slightly reduced if this shift is assumed.}

\subsection{Implications for Quantum Gravity Models}

An  effective  field theory  approximation  for  possible LIV  effects
induced by Planck-scale suppressed  quantum gravity for $E \ll M_{Pl}$
was considered in Ref.  ~\cite{ma09}.  These authors explored the case
where a  perturbation to  the energy-momentum dispersion  relation for
free particles would be produced  by a CPT-even dimension six operator
suppressed  by a  term  proportional to  $M_{Pl}^{-2}$. The  resulting
dispersion relation for a particle of type $a$ is

\begin{equation}
E_a^2 = p_a^2 + m_a^2 + \eta_a \left( {{p^4}\over{M_{Pl}^2}} \right)
\label{QG}
\end{equation} 

In order  to explore the  implications of our constraints  for quantum
gravity,  one  can  take  the  perturbative terms  in  the  dispersion
relations for both protons and pions, to be given by the dimension six
dispersion   terms  in   equation  (\ref{QG})   above.    Making  this
identification, the LIV constraint of $\delta_{\pi p} < 4.5
\times 10^{-23}$ in  the  fiducial  energy  range around  $E_f  =  100$  EeV
indirectly implies  a powerful limit on the  representation of quantum
gravity  effects in an  effective field  theory formalism  with Planck
suppressed dimension  six operators.  Equating  the perturbative terms
in  both the  proton and  pion dispersion  relations, one  obtains the
relation~\cite{st09}

\begin{equation}
2\delta_{\pi p} \simeq (\eta_{\pi} - 25 \eta_{p})
\left({{0.2E_f}\over{M_{Pl}}}\right)^2 ,
\label{dim6}
\end{equation}

\noindent where the pion fiducial energy is taken to to be $\sim 0.2 E_f$,  
as at  the $\Delta$ resonance that dominates  photopion production and
the GZK effect~\cite{st68}.   Equation (\ref{dim6}), together with the
constraint $\delta_{\pi p} <  4.5 \times 10^{-23}$, indicates that any
LIV from dimension six operators is suppressed by a factor of at least
${\cal{O}}(10^{-6}  M_{Pl}^{-2})$, except  in the  unlikely  case that
$\eta_{\pi}- 25  \eta_{p} \simeq 0$.   These results are  in agreement
with those obtained  independently by Maccione et al.   from the Monte
Carlo runs~\cite{ma09}.   It can thus  be concluded that  an effective
field  theory representation  of  quantum gravity  with dimension  six
operators  that suppresses  LIV by  only a  factor of  $M_{Pl}^2$ {\it
i.e.}  $\eta_p,  \eta_{\pi} \sim 1$,  is effectively ruled out  by the
UHECR observations.

\section{Beyond Constraints: Seeking LIV}

As we have seen (see Figure  \ref{Auger}), even a very small amount of
LIV that  is consistent with  both a GZK  effect and with  the present
UHECR data can lead to a ``recovery'' of the primary UHECR spectrum at
higher energies. This is the  clearest and the most sensitive evidence
of an LIV signature. The  ``recovery'' effect has also been deduced in
Refs.~\cite{ma09} and ~\cite{bi09}
\footnote{In Ref.~\cite{bi09},  a recovery effect is  also claimed for
high proton energies in the  case when $\delta_{\pi p} < 0$.  However,
we have noted that  the `quasi-vacuum \v{C}erenkov radiation' of pions
by  protons in  this case  will  cut off  the proton  spectrum and  no
``recovery'' effect will  occur.}. In order to find  it (if it exists)
three conditions must exist: ({\it i}) sensitive enough detectors need
to be built, ({\it ii}) a  primary UHECR spectrum that extends to high
enough energies ($\sim$ 1000 EeV) must exist, and ({\it iii}) one much
be able to distinguish the LIV signature from other possible effects.

\subsection{Obtaining UHECR Data at Higher Energies}

We now  turn to examining the  various techniques that can  be used in
the  future  in  order  to  look  for a  signal  of  LIV  using  UHECR
observations.   As   can  be  seen  from   the  preceding  discussion,
observations of higher energy  UHECRs with much better statistics than
presently obtained  are needed in order  to search for  the effects of
miniscule Lorentz invariance violation on the UHECR spectrum.

\subsubsection{Auger North}

Such an increased  number of events may be  obtained using much larger
ground-based  detector  arrays.   The  {\it Auger}  collaboration  has
proposed to build  an ``{\it Auger North''} array  that would be seven
times larger  than the present  southern hemisphere Auger  array ({\tt
http://www.augernorth.org}).

\subsubsection{Space Based Detectors}

Further  into  the future,  space-based  telescopes  designed to  look
downward  at large  areas of  the  Earth's atmosphere  as a  sensitive
detector  system  for giant  air-showers  caused  by trans-GZK  cosmic
rays. We  look forward to  these developments that may  have important
implications for fundamental high energy physics.

Two potential  spaced-based missions have been proposed  to extend our
knowledge of  UHECRs to higher  energies.  One is {\it  JEM-EUSO} (the
Extreme  Universe  Space  Observatory) ~\cite{EUSO},  a  one-satellite
telescope  mission proposed to  be placed  on the  Japanese Experiment
Module (JEM)  on the International  Space Station.  The other  is {\it
OWL}  (Orbiting  Wide-angle   Light  Collectors)  ~\cite{OWL},  a  two
satellite mission for stereo viewing, proposed for a future free-flyer
mission.   Such  orbiting  space-based  telescopes with  UV  sensitive
cameras will have  wide fields-of-view (FOVs) in order  to observe and
use large  volumes of  the Earth's atmosphere  as a  detecting medium.
They will thus trace the atmospheric fluorescence trails of numbers of
giant  air  showers  produced  by  ultrahigh energy  cosmic  rays  and
neutrinos.   Their large  FOVs will  allow the  detection of  the rare
giant air  showers with energies higher than  those presently observed
by  ground-based detectors such  as {\it  Auger}.  Such  missions will
thus  potentially open  up  a new  window  on physics  at the  highest
possible observed energies.

\section{Conclusions}

The   {\it  Fermi}  timing   results  for   GRB090510  rule   out  and
string-inspired  D-brane model  predictions as  well as  other quantum
gravity predictions of a retardation of photon velocity that is simply
proportional  to  $E/M_{QG}$  because  they would  require  $M_{QG}  >
M_{Pl}$.  More indirect results  from \gray\ birefringence limits, the
non-decay of 50 TeV \grays\ from  the Crab Nebula, and the TeV spectra
of  nearby AGNs  also place  severe  limits on  violations of  special
relativity  (LIV).   Limits   on  Lorentz  invariance  violation  from
observations   of   ultrahigh   energy  cosmic-rays   provide   severe
constraints for  other quantum gravity  models, appearing to  rule out
retardation  that is  simply proportional  to  $(E/M_{QG})^2$. Various
effective field theory frameworks lead to such energy dependences.

New  theoretical models of  Planck scale  physics and  quantum gravity
need  to  meet all  of  the  present  observational constraints.   One
scenario  that may  be considered  is that  gravity, {\it  i.e.}  $G$,
becomes weaker  at high energies.  We  know that the  strong, weak and
electromagnetic interactions all have energy dependences, given by the
running  of  the  coupling  constants.   If $G$  decreases,  then  the
effective $\lambda_{Pl}  = \sqrt{G\hbar /c^3}$ would  decrease and the
effective $M_{Pl} = \hbar  / (\lambda_{Pl}c)$ would increase.  In that
case,  the space-time  quantum  scale  would be  less  than the  usual
definition  of  $\lambda_{Pl}$.  Such  speculation  is presently  {\it
cogitare ex arcis}, but might be  plausible if a transition to a phase
where  the various  forces are  unified occurs  at very  high energies
~\cite{st80}.

At  the  time  of   the  present  writing,  high  energy  astrophysics
observations  have led to  strong constraints  on LIV.   Currently, we
have no positive  evidence for LIV. This fact,  in itself, should help
guide theoretical research on quantum gravity, already ruling out some
models.   Will  this   lead  to  a  new  null   result  comparable  to
Michelson-Morley?  Will a totally  new concept  be needed  to describe
physics at the Planck scale? If all of the known forces are unified at
the Planck scale, this would not be surprising. One thing is clear:
a consideration of all empirical data will be necessary in order to
finally arrive at a true theory of physics at the Planck scale.

\end{document}